# A COMPUTER VIRUS PROPAGATION MODEL USING DELAY DIFFERENTIAL EQUATIONS WITH PROBABILISTIC CONTAGION AND IMMUNITY.

M. S. S. Khan

Department of Computer Science ,College of Information & Computer Technology
Sullivan University, Louisville, KY 40205.

*ABSTRACT*

*The SIR model is used extensively in the field of epidemiology, in particular, for the analysis of communal diseases. One problem with SIR and other existing models is that they are tailored to random or Erdos type networks since they do not consider the varying probabilities of infection or immunity per node. In this paper, we present the application and the simulation results of the pSEIRS model that takes into account the probabilities, and is thus suitable for more realistic scale free networks. In the pSEIRS model, the death rate and the excess death rate are constant for infective nodes. Latent and immune periods are assumed to be constant and the infection rate is assumed to be proportional to $I(t)/N(t)$, where N (t) is the size of the total population and I(t) is the size of the infected population. A node recovers from an infection temporarily with a probability p and dies from the infection with probability (1-p).*

## 1. INTRODUCTION

The growth of the internet has created several challenges and one of these challenges is cyber security. A reliable cyber defense system is therefore needed to safeguard the valuable information stored on a system and the information in transit. To achieve this goal, it becomes essential to understand and study the nature of the various forms of malicious entities such as viruses, trojans, and worms, and to do so on a wider scale. It also becomes essential to understand how they *spread* throughout computer networks.

A computer virus is a malicious computer code that can be of several types, such as a trojan, worm, and so on [1, 14]. Although each type of malicious entity has a different way of spreading over the network, they all have common properties such as infectivity, invisibility, latency, destructibility and unpredictability [10].

More recently, we have also started witnessing viruses that can spread on social networks. These viruses spread by infecting the accounts of social network users, who click on any option that may trigger the virus's takeover of the user's sharing capabilities, resulting in the spreading of these malicious programs without the knowledge of the user. While their inner coding might be different and while their triggering and spreading mechanisms (on physical vs. virtual social networks) may be different, both traditional viruses such as worms and social network viruses share, in common, the critical property of proliferation via spreading through a network (physical or social).

DOI : 10.5121/ijcnc.2014.6508      111

These malicious programs behave similarly to an infection in a human population. This, in turn, allows us to draw comparisons between the study of epidemiology, in particular the mathematical aspect of infectious diseases[2] and the behavior of a computer virus in a computer network. This is generally studied via mathematical models of the spread of a virus or disease. Any mathematical model's ability to mimic the behavior of the infection largely depends on the assumptions made during the modeling process.

Most of the existing epidemiology models are modified versions of the classical Kermack and McKendrick's [15] model, more commonly known as the SIR (Susceptible/Infected/Recovered) model. For example, Hethcote [9] proposed a version of the SIR model, in which it was assumed that the total population was constant. But in real world scenarios, the population will change in time. Thus, this model was later improved by Diekmann and Heersterbeek[8] by assuming that:

I. The population size changes according to an exponential demographic trend.
II. The infected individuals cannot reproduce.
III. The individuals acquire permanent immunity to further infection when removed from the infected class.

The eigenvalue approach has recently emerged as one of the popular techniques to analyze virus or malicious entity propagation in a computer network. Wang et al. [17] have associated the epidemic threshold parameter for a network with the largest eigenvalue of its adjacency matrix. This technique works only with the assumption that the eigenvalues exist. There are also certain restrictions on the size of the adjacency matrix, since calculating eigenvalues may not be easy or even possible for large matrices.

In this paper, we apply the malicious object transmission model [12] in complete and scale free networks, that assumes a variable population and with constant latent and immune periods. The current model extends the classical SIR model proposed in [11] to a *probabilistic* SEIRS [12] type model in several directions:

I. It includes an *Exposed* class in addition to the *Susceptible*, *Infected* and *Recovered* classes.
II. It furthermore includes a constant exposition period $\omega$ and a constant latency period $\tau$.
III. When a node is removed from the infected class, it recovers with a temporary immunity with a probability *p* and dies due to the attack of the malicious object with probability (1-*p*).

The rest of this paper is organized as follows. In section 2, we introduce the classical SIR model and implement it on a small network. After presenting simulation and statistical results, we conclude that the SIR model does not capture the realistic nature of the virus propagation in a computer network. Thus, we consider the more realistic scale free network in section 3 and give some historical background about such networks. A system of delay differential equations is then presented that includes a probabilistic temporary immunity with probability *p*. A complete mathematical justification follows the introduction of *p*SEIRS model. Section 4 discusses the mathematical conditions for an infection free equilibrium. We also discuss the stability analysis of the proposed model using the phase plane plots. We then present simulation experiments to validate the scalability aspect of the proposed model and compare it with the classical SEIRS model. Section 5 presents our conclusion, then discusses some limitations of compartment type mathematical models, and finally closes with some possible directions for future work in this area.

## 2. The Classical SIR Model

The SIR (Susceptible/Infected/Recovered) model [15] is used extensively in the field of epidemiology, in particular, for the analysis of communal diseases, which spread from an infected individual to a population. So it is natural to divide the population into the separate groups of susceptible (S), infected (I) and those who have recovered or become immune to a type of infection or a disease (R). These subdivisions of the population, which are also different stages in the epidemic cycle, are sometimes called *compartments*. A total population of size *N* is thus divided into three stages or compartments:

$$N = S + I + R.$$

Since the population is going to change with time (*t*), we have to express these stages as a function of time, i.e., *S* (*t*), *I* (*t*) and *R* (*t*). Here $\beta$ (infection rate) and $\alpha$ (recovery rate) are the transition rates between $S \to I$ and $I \to R$, respectively, and are both in [0, 1].

The underlying assumption in the traditional SIR model is that the population is homogenous; that is, everybody makes contact with each other randomly. So from a graph theoretical point of view, this population represents a complete graph, which would be the case of a computer network that is totally connected. In this situation, the classical SIR model may be sufficient enough to understand the malicious object's propagation. The dynamics of the epidemic evolution can be captured with the following homogenous differential equations:

$$\frac{dS}{dt} = -\beta IS$$
$$\frac{dI}{dt} = \beta IS - \alpha I \qquad (1)$$
$$\frac{dR}{dt} = \alpha I \ .$$

Based on (1), we define a basic reproduction rate as

$$R_0 = \frac{\beta}{\alpha}$$

$R_0$ is a useful parameter to determine if the infection will spread through the population or not. That is, if $R_0 > 1$ and $I(0) > 1$, the infection will be able to spread in a population, to a stage that is called *Endemic Equilibrium* and is given by

$$\lim_{t \to \infty}(S(t), I(t), R(t)) \to \left(\frac{N}{R_0}, \frac{N}{\beta}(R_0 - 1), \frac{\alpha N}{\beta}(R_0 - 1)\right). \qquad (2)$$

If $R_0 < 1$, infection will die out in the long run. This is called *disease free equilibrium* and is represented as

$$\lim_{t \to \infty}(S(t), I(t), R(t)) \to (N, 0, 0) \ . \qquad (3)$$

### 2.1: Simulation Results of the SIR Model with A Small Complete Network

The simulation results of the SIR model in a $K_{12}$ network (complete graph with 12 nodes) are presented in Fig. 1(a) and 1 (b) with different values of $\beta$ and $\alpha$.

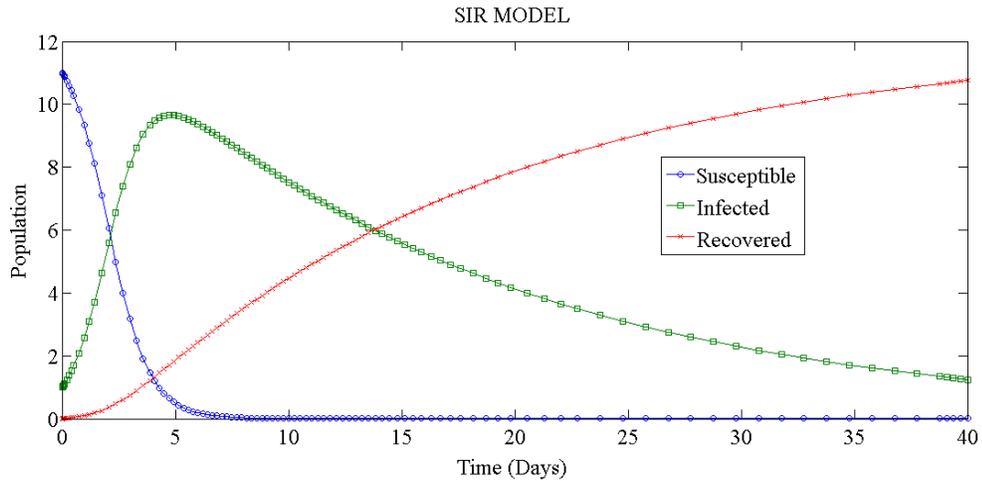

Fig.1 (a): SIR model based on $\beta = 0.06$, $\alpha = 0.1$

In Fig 1(a), $R_0 = 0.6 < 1$. So, the infection will not spread through the entire network.

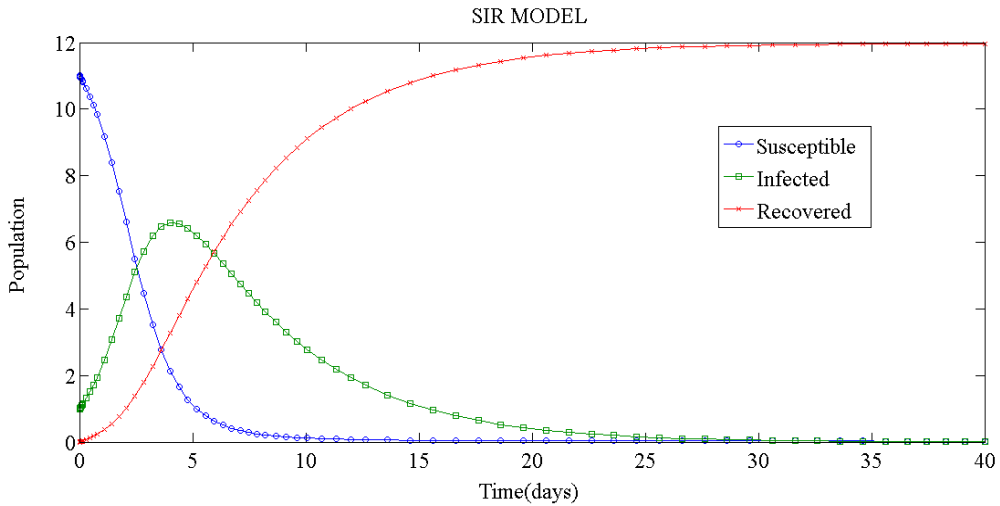

Fig. 1 (b): SIR model based on $\beta = 0.1$, $\alpha = 0.2$

In Fig. 1 (b), $R_0 = 2 > 1$. In this case, infection will spread to most of the nodes and the network will be in an endemic stage.

We gleam useful information about the virus propagation from Fig. 1(a) and 1(b) in a $K_{12}$ type network and make an analysis based on these simulations. Two obvious observations are the long-term behavior of the infection and the speed at which the population gets infected. Also, we can determine the rate at which the population is getting immune to the infection, i.e., the removal rate from the susceptible group. Information such as the maximum, minimum, and mean rate of the infected population can also be determined in each class. A statistical analysis of the cases in Fig. 1(a) and 1(b) are presented in Table 1 and Table 2.

| Table 1: Statistical Analysis of Fig. 1(a) | | | | Table 2: Statistical Analysis of Fig. 1(b) | | | |
|---|---|---|---|---|---|---|---|
|  | Susceptible | Infected | Recovered |  | Susceptible | Infected | Recovered |
| Min | 0.008 | 0.1297 | 0 | Min | 0 | 0.090 | 0 |
| Max | 11 | 7.189 | 11 | Max | 11 | 9.954 | 11 |
| Mean | 3.619 | 2.522 | 5.859 | Mean | 1.965 | 4.832 | 5.203 |

This analysis is based on some crucial assumptions, such as:

- Once a node is removed or recovered from the infection, it develops immunity and it cannot be infected again.
- The population is homogenous and well mixed.
- There is no latency, i.e., there is no time delay between exposure to the infection and actually getting infected.

The above assumptions make it clear that in order to capture the essence of virus propagation in a computer network, in a more realistic sense, the SIR model is not sufficient to study virus propagation. In order to address this limitation, we must assume a network with a scale free distribution of the node connectivity degrees. We also must take into account the real phenomena of possible re-infection and the time delay incurred during incubation (between exposition and infection). For these reasons, in the following sections, we will first briefly review scale free networks and then present the $p$SEIRS[12] model that addresses the limitations of the classical model.

## 3. Application of Probabilistic SEIRS Model

### 3. 1: Scale Free Networks

Scale free networks were first observed by Derek de Solla Price in 1965, when he noticed a heavy-tailed distribution following a Pareto distribution or Power law in his analysis of the network of citations between scientific papers. Later, in 1999, the work of Barabasi and Albert [3] reignited interest in scale free networks.

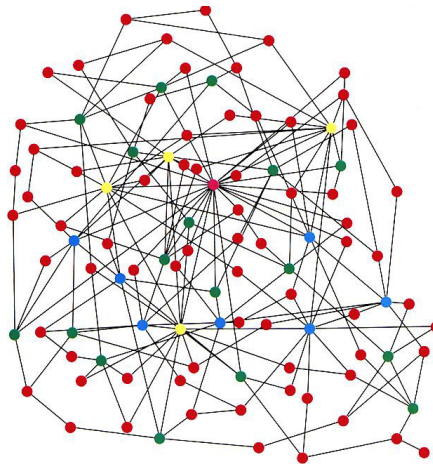

Fig. 2: Example of a scale free network.

For example, Lloyd and May [11] generated the scale free network, depicted in Fig. 2, using the network generating algorithm of Barabasi and Albert [3]. There are 110 nodes that are colored according to their connectivity degree in red, green, blue, and yellow, with the most highly connected nodes colored in blue. It is worth noting that there are only a few highly connected nodes, while the majority of the nodes have only a few connections. Thus, in a classical scale free network, the connectivity of a node follows a power law distribution [11].

### 3.2: Probabilistic SEIRS Model: *p*SEIRS

In the time delay model, parameter, $\gamma$, represents the probability of spreading the infection in one contact. It is obvious that the rate of propagation is proportional to the connectivity of the node. The propagation rate does not change in time throughout the network.

In order to capture more realistic dynamics within real world scale free networks, several more variables compared to the classical SIR model are used and to capture more realistic behavior, delay is consider on the network. An additional stage (*Exposed*) in the model represents the phenomenon of incubation, leading to a delay between susceptibility to infection and actual infection. The Exposed stage makes the model a SEIR model instead of a SIR model. Furthermore, the *p*SEIRS model is different from the classical SEIRS model (the latter is obtained for an immunity probability *p* = 1). The resulting model is described in terms of the following variables and constants:

$N(t)$: Total Population size
$S(t)$: Susceptible Population
$E(t)$: Exposed Population
$I(t)$: Infected Population
$R(t)$: Recovered Population
$\beta$ : Birth rate.
$\mu$ : Death rate due to causes other than an infection by a virus.
$\varepsilon$ : Death rate due an infection by a virus, it is constant.
$\alpha$ : Recovery rate which is constant.
$\gamma$ : Average number of contacts of a node, also equal to the probability of spreading the virus in one contact.
$\omega$ : Latency period or time delay, which is a constant i.e., the time between the exposed and infected stages.
$\tau$ : Period of temporary immunity, which is a positive constant.
$p$ : Probability of temporary immunity of a node after recovery.

Once an infection is introduced into a network, its nodes will become *susceptible* to the infection and, in due course, will get *infected*. Once a node is *exposed*, an incubation period is observed, which is captured by the new time delay parameter, which therefore models reality better: any infection goes through an incubation period before it propagates. After infection, anti-virus software may be executed to treat an infected node, thus providing it with *temporary immunity*. It is important to realize that there is no permanent immunity in a real network, thus an immune node may revert to the *susceptible* stage again. All these stages of infection from *susceptible* to *recovered*, and the phases in between, are shown in Fig. 3.

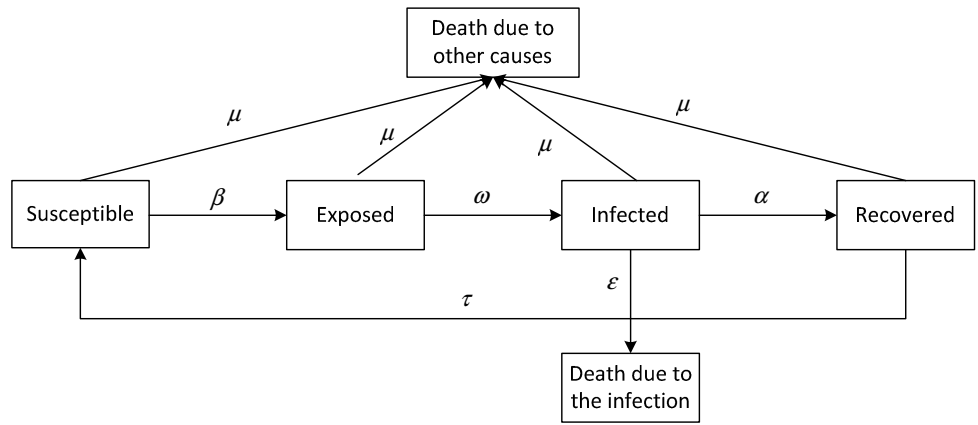

Fig. 3: Flow of infection in a SEIRS model.

The *p*SEIRS model in Fig. 3 works under the following assumptions:

- Any new node entering the network is susceptible.
- The death rate of a node, when the death is due to other reasons that are separate from virus infection, is constant throughout the network.
- The death rate of a node due to infection is also constant.
- The latency period and immune period are constant.
- The incubation period for the exposed, infected and recovered nodes is exponentially distributed.
- Once infected, a node can either (i) recover and become immune with probability of temporary immunity *p*, or (ii) die from the infection with probability (1-*p*).

Once a network is attacked (some initial nodes become infective), any node that is in contact with the infective nodes becomes exposed, i.e., infected but not infectious. Such nodes remain in the incubation period before becoming infective and have a constant period of temporary immunity, once an effective anti-virus treatment is run. The total population size is expressed as

$$N(t) = S(t) + I(t) + E(t) + R(t) \,. \tag{4}$$

Also, it is important to understand that the infection remains in the network for at least *κ = max (τ, ω)*, so that we have an initial perturbation period. The model in Fig. 3, has the following form for *t > τ*:

$$\frac{dS(t)}{dt} = \beta N(t) - \mu S(t) - \gamma \frac{S(t)I(t)}{N(t)} + \alpha I(t-\tau)e^{-\mu\tau} \tag{5}$$

$$E(t) = \int_{t-\omega}^{t} \gamma \frac{S(x)I(x)}{N(x)} e^{-\mu(t-x)} dx, \tag{6}$$

$$\frac{dI(t)}{dt} = \gamma \frac{S(t-\omega)I(t-\omega)}{N(t-\omega)} e^{-\mu\omega} - (\mu + \varepsilon + \alpha)I(t), \tag{7}$$

$$R(t) = \int_{t-\tau}^{t} p\alpha I(x) e^{-\mu(t-x)} dx. \tag{8}$$

equations (5) - (8) are called an integro-differential equation system. On differentiating (6) and (8), we get the following:

$$\frac{dE(t)}{dt} = \gamma \frac{S(t)I(t)}{N(t)} - \gamma \frac{S(t-\omega)I(t-\omega)}{N(t-\omega)} e^{-\mu\omega} - \mu E(t), \tag{9}$$

$$\frac{dR(t)}{dt} = p\alpha I(t) - \alpha I(t-\tau)e^{-\mu\tau} - \mu R(t). \tag{10}$$

The system of differential equations (5), (7), (8) and (10) is collectively called a differential difference equation system and we will refer to it as M1. That is

$$\left.\begin{aligned}
\frac{dS(t)}{dt} &= \beta N(t) - \mu S(t) - \gamma \frac{S(t)I(t)}{N(t)} + \alpha I(t-\tau)e^{-\mu\tau} \\
\frac{dE(t)}{dt} &= \gamma \frac{S(t)I(t)}{N(t)} - \gamma \frac{S(t-\omega)I(t-\omega)}{N(t-\omega)} e^{-\mu\omega} - \mu E(t) \\
\frac{dI(t)}{dt} &= \gamma \frac{S(t-\omega)I(t-\omega)}{N(t-\omega)} e^{-\mu\omega} - (\mu + \varepsilon + \alpha) I(t) \\
\frac{dR(t)}{dt} &= p\alpha I(t) - \alpha I(t-\tau)e^{-\mu\tau} - \mu R(t).
\end{aligned}\right\} M1$$

It is important for the continuity of the solution to M1 to have

$$E(0) = \int_{-\omega}^{0} \frac{\gamma S(x)I(x)}{N(x)} e^{\mu x} \, dx \tag{11}$$

$$R(0) = \int_{-\tau}^{0} p\alpha I(x) e^{\mu x} \, dx. \tag{12}$$

It is worth noting that M1 is different from the model by Cooke and Driessche [5] because they do not discuss the probability of immunity when a node recovers from the infection. M1, on the other hand, is a probabilistic model; here, a node acquires a temporary immunity with probability *p* and dies with probability 1-*p*. M1 is also different from Yan and Liu [16] because their model assumes that a recovered node acquires *permanent* immunity, which may not be the case in real networks.

**Theorem 1[15]:** A solution of the integro-differential equations system (5-8) with *N (t)* given by (4) satisfies (9) and (10). Conversely, let *S (t), E (t), I (t)* and *R (t)* be a solution of the delay differential equation system (M1), with *N (t)* given by (4) and the initial condition on $[-\tau, 0]$. If in addition,

$$E(0) = \int_{-\omega}^{0} \gamma S(x)I(x)e^{\mu x} dx \quad \text{and} \quad R(0) = \int_{-\tau}^{0} p\alpha I(x)e^{\mu x} dx,$$

then this solution satisfies the integro-differential system in (5-8).

## 4. Infection Free Equilibrium

Let D be a close and bounded region such that

$$D = ((S,E,I,R): S,E,I,R \geq 0, S+E+I+R = 1). \tag{13}$$

In (13), we consider the equilibrium of M1, when the infection $I = 0$, then $E = R = 0$ and $S = 1$. This is the only equilibrium on the boundary of $D$ and we also have the threshold parameter for the existence of the interior equilibrium:

$$R_0 = \frac{\gamma e^{-\beta\omega}}{\varepsilon + \beta + \alpha} . \tag{14}$$

The threshold parameter $R_0$ is a measure of the strength of the infection. The quantity $\frac{1}{\beta+\varepsilon+\alpha}$ is the mean waiting time in the infection class, thus $R_0$ is the mean number of contacts of an infection during the mean time in the infection class. Asymptotically, as time $t \to \infty$, we get an infection free equilibrium [5].

The system of equations in M1 will always reach a disease free equilibrium if $R_0 < 1$, thus all the solutions starting in $D$ will approach the disease free equilibrium. On the other hand, when $R_0 > 1$, we have an endemic equilibrium meaning that the infection will not die off in the long run.

### 4.1. Simulation Results with the *p*SEIRS Model

The dynamic behavior of the system M1 is presented in Fig. 4, with the following values:

$\beta = 0.330, \mu = 0.006, \varepsilon = 0.060, \alpha = 0.040, \gamma = 0.308, \omega = 0.15, \tau = 30, p = 1.$

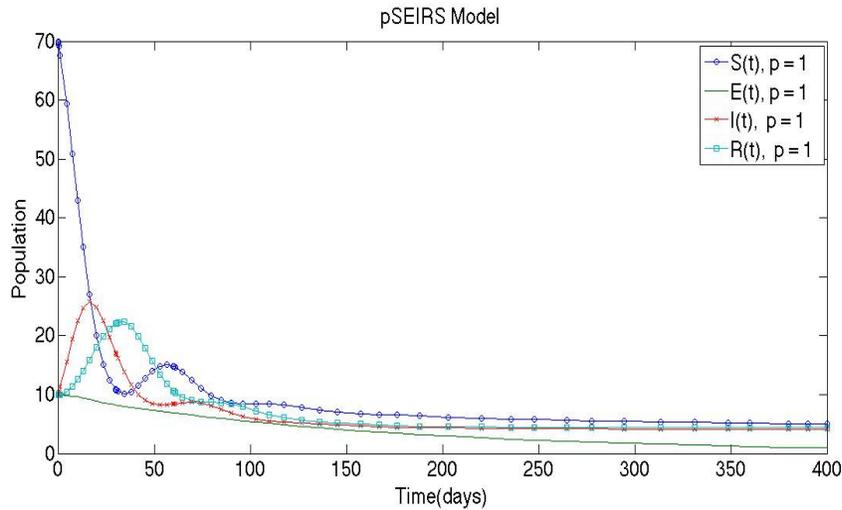

Fig. 4: *p*SEIRS model.

Under the above assumption, we find that the threshold parameter is $R_0 = 7.77$. Hence, the system is in the endemic state. We also have *p*=1, thus assuming that all the recovered nodes have permanent immunity. In Fig. 5, we present the overall impact on the population with the above-mentioned parameters.

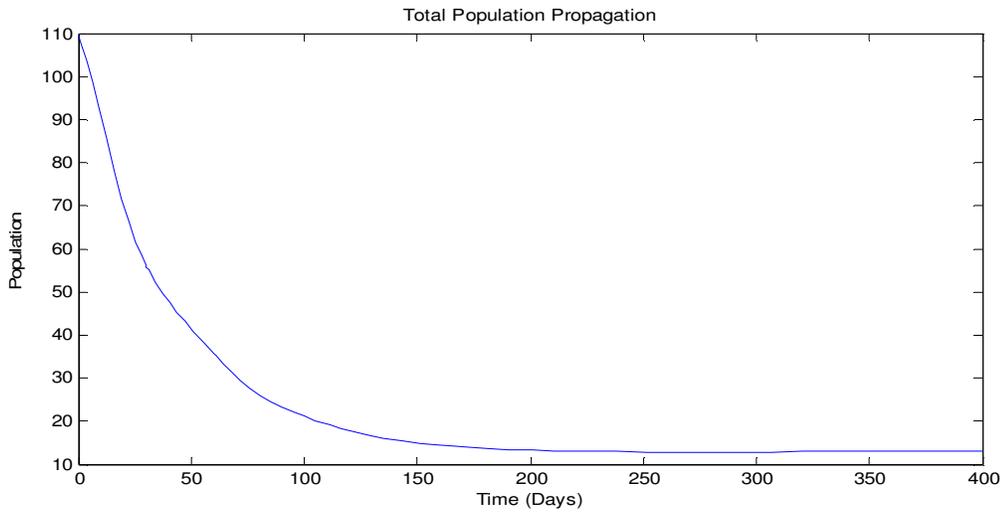

Fig. 5: Population propagation under the *pSEIRS* model with *p* =1.

We notice from Table 3, that there are 33 nodes, which will get infected, and that the mean rate of infection is 12.37 nodes, while the number of recovered nodes is 20.00 with a mean recovery rate of 7.5 nodes

Table 3: Statistical Analysis of Fig. 4.

|  | Susceptible | Exposed | Infected | Recovered |
|---|---|---|---|---|
| Min | 4.4 | 1.64 | 4.1 | 1.027 |
| Max | 70 | 10 | 33.46 | 19.94 |
| Mean | 16.59 | 6.16 | 12.37 | 7.5 |

Stability is an important issue with any dynamical system. In Fig. 6, we present the phase plane portrait of our proposed model, which shows that we have an endemic equilibrium; hence our modeled system is asymptotically stable at the equilibrium (4.511, 0.9362, 4.161).

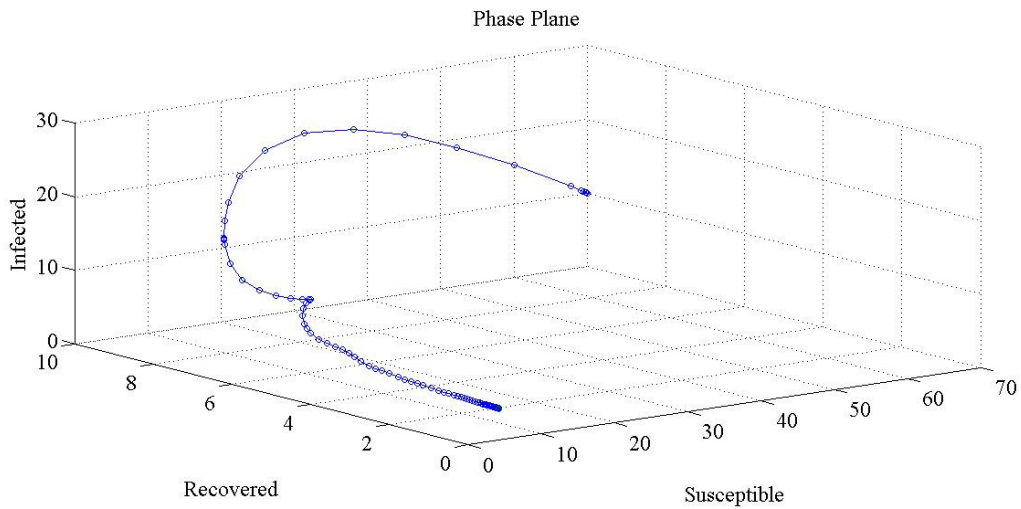

Fig. 6: Phase Plan Portrait for temporary immunity probability *p* = 1 and latency period $\omega = 0.15$.

We now look at the dynamics with the temporary immunity probability *p* = 0.4. We can clearly notice that the recovery *R* (t) has declined in Fig. 7 compared to Fig. 4, since we have chosen a

120

International Journal of Computer Networks & Communications (IJCNC) Vol.6, No.5, September 2014

lower probability for temporary immunity.

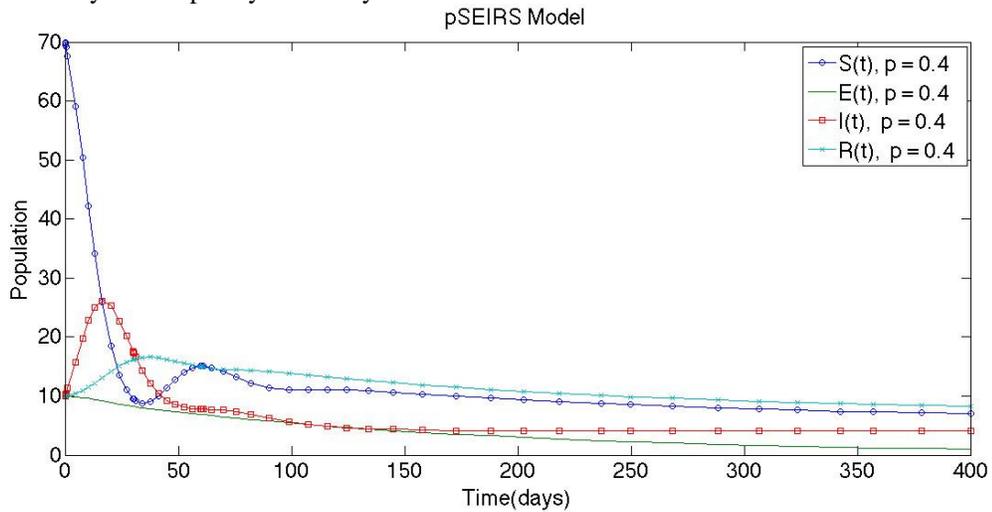

Fig. 7: *p*SEIRS model.

We also notice from Table 4, that 34 nodes will get infected and the mean rate of infection is 13.00 nodes while the number of recovered nodes is 14.00, with a mean recovery rate of 7.255 node.

Table 4: Statistical Analysis of Fig. 7.

|  | Susceptible | Exposed | Infected | Recovered |
|---|---|---|---|---|
| Min | 4.216 | 1.66 | 4.283 | 2.027 |
| Max | 70 | 10 | 33.94 | 13.94 |
| Mean | 17.71 | 6.405 | 13.19 | 7.255 |

Because of the lower temporary immunity rate, we expected the population to decline further, and this is confirmed in Fig. 8.

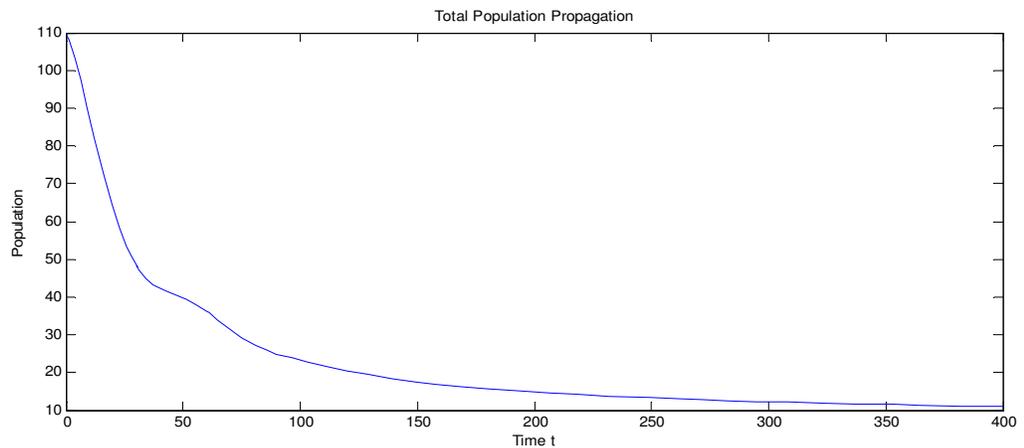

Fig. 8: Population propagation under the *p*SEIRS model with *p* = 0.4.

In Fig. 9, we again see an asymptotically stable system at the equilibrium (7.054, 0.9407, 4.05).

121

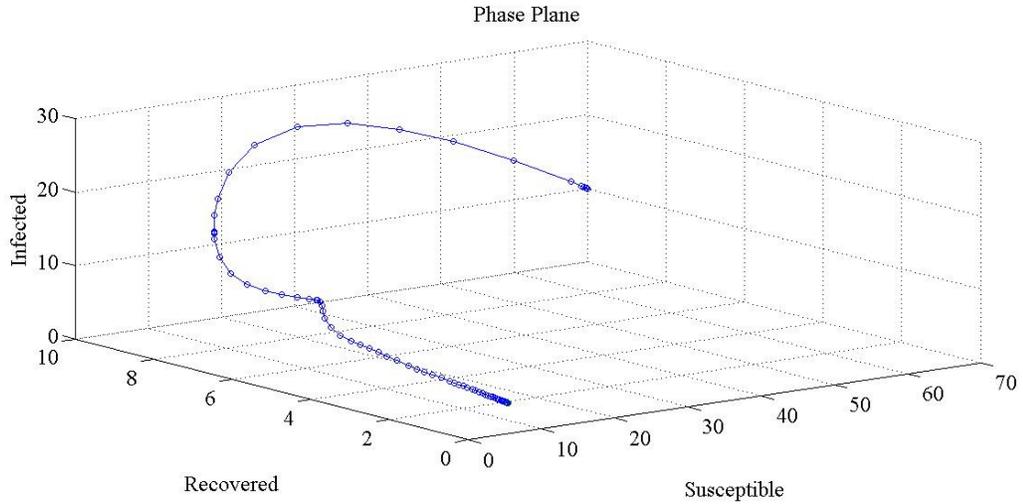

Fig. 9: Phase Plane Portrait for temporary immunity probability $p = 0.4$ and latency period $\omega = 0.15$.

Now consider the case, when the latency period is increased. We also consider the temporary immunity probability to be $p = 1$ and all the other parameters remain the same as before. The latency effect on the virus propagation dynamics can be clearly seen in Fig. 10.

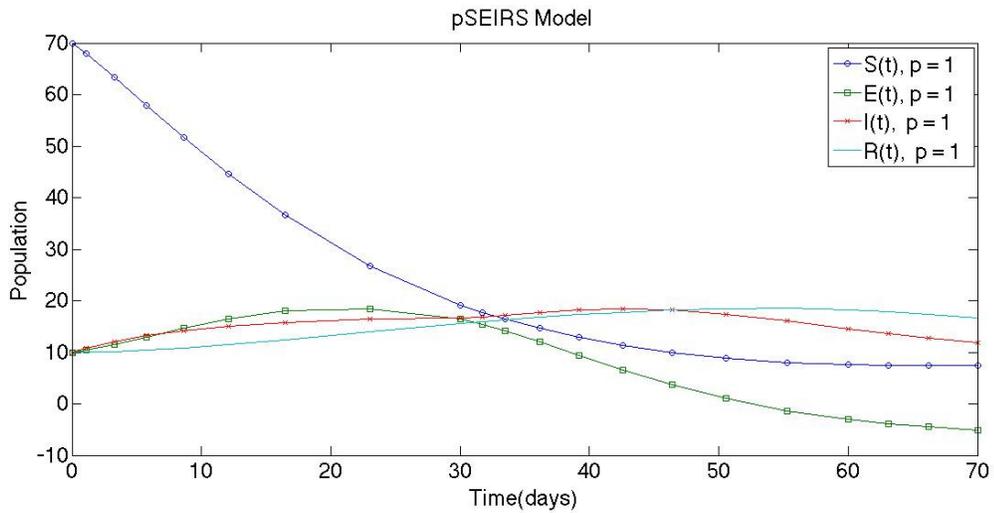

Fig. 10: $p$SEIRS model with $\omega = 30$.

Infection is less likely to become an endemic in this case because $R_0 = 0.3703 < 1$. Instead, we expect to have an asymptotically stable system at a disease free equilibrium (1.118e-006, 20.63, 11.65), as shown in Fig. 11.

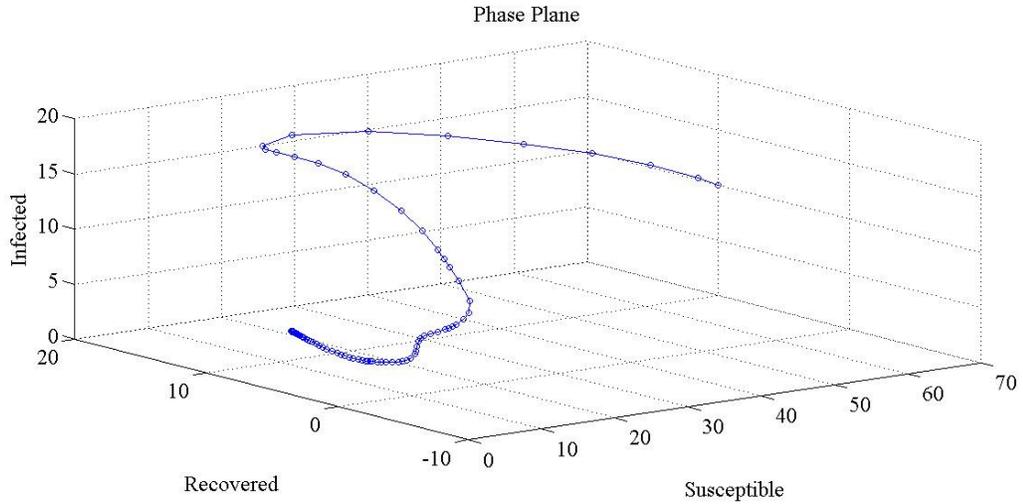

Fig. 11: Phase Plane Portrait for temporary immunity probability $p = 1$ and a longer latency period ($\omega = 30$.)

### 4.2. Scalability of the pSEIRS model

So far, we have used relatively speaking a small network of 110 nodes in a scale free network. We now consider a larger network of 5000 nodes, as shown in Fig. 12. The network was generated in the Matlab environment using the Barabasi-Albert graph generation algorithm [3].

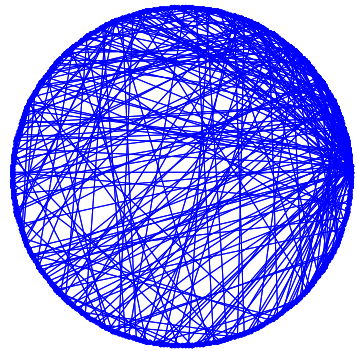

Fig. 12: Scale free network of 5000 nodes.

It is important to realize that it is very typical for most of the networks arising from social networks, World Wide Web links, biological networks, computer networks and many other phenomena, to follow a power law in their connectivity. Thus, networks are conjectured to be scale free[4]. Here, the immunity probability is $p = 0.5$ and the rest of the parameters remain the same as before. So, under these assumptions, $R_0 = 8.621329079589127e\text{-}001 < 1$, meaning that the infection is less likely to become endemic. The global behavior of the proposed model is presented in Fig. 13.

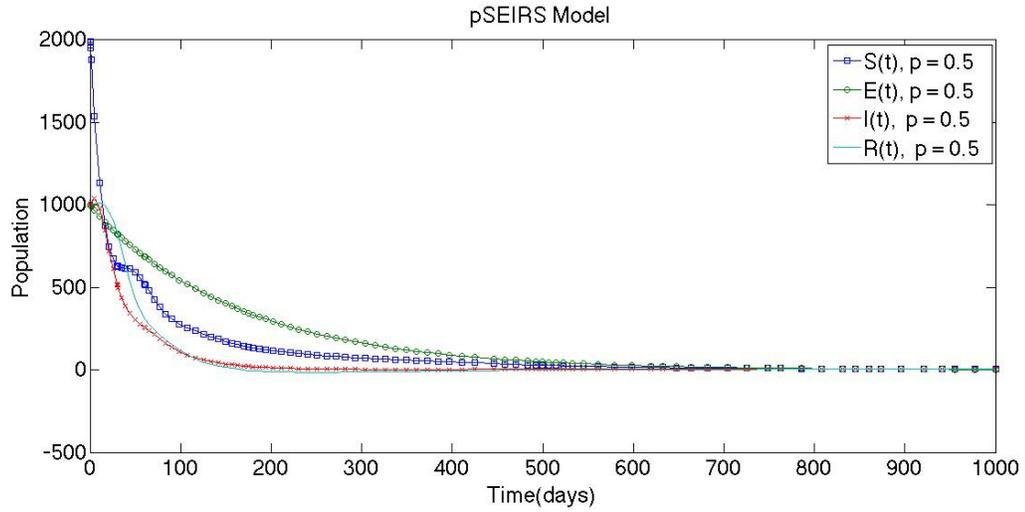

Fig. 13: *p*SEIRS Model with 5000 nodes.

Moreover, we have an asymptotically stable system at a disease free equilibrium of (5.778, 2.51, 4.158). This can be seen in Fig. 14.

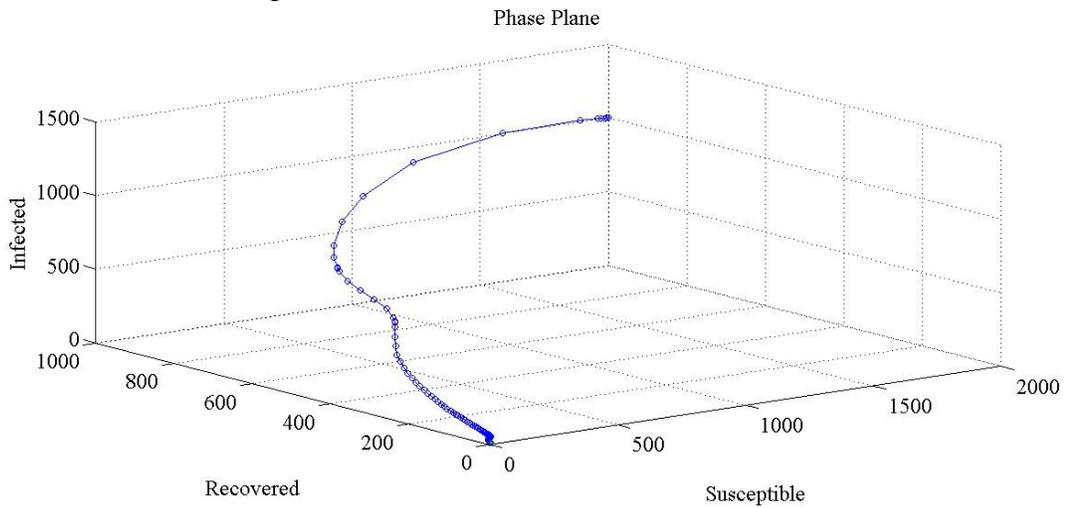

Fig. 14: Phase Plane Portrait.

The overall population propagation is presented in Fig. 15, with $p = 0.5$.

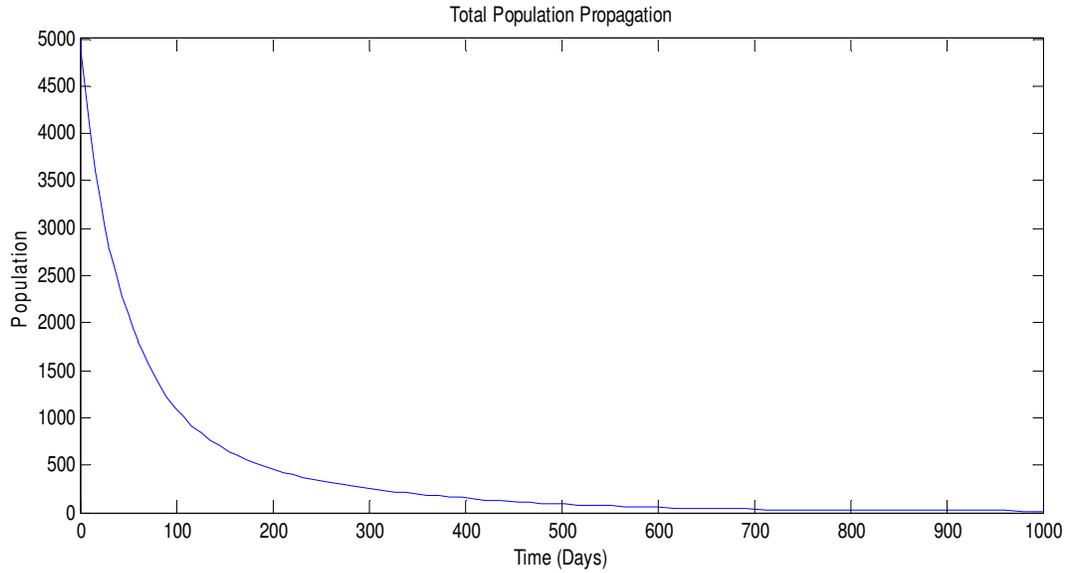

Fig. 15: Population propagation under the *p*SEIRS model with *p* = 0.5 and 5000 nodes.

We now compare the *p*SEIRS model and the classical SEIRS model (the latter is obtained for an immunity probability *p* = 1). In Fig. 16 and Fig. 17, we can clearly see the difference between the recovery rates *R*(t) for the different models.

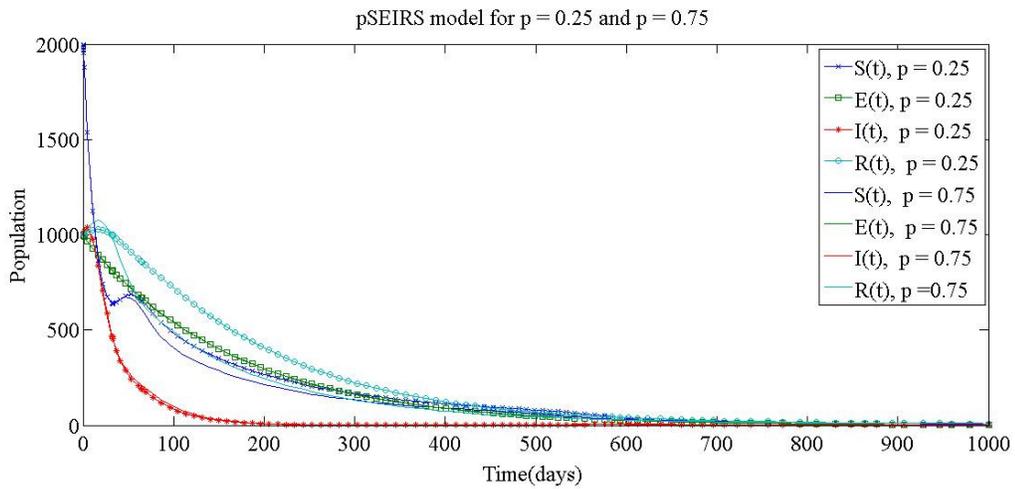

Fig. 16: *p*SEIRS model with *p* = 0.25 and *p* =0.75.

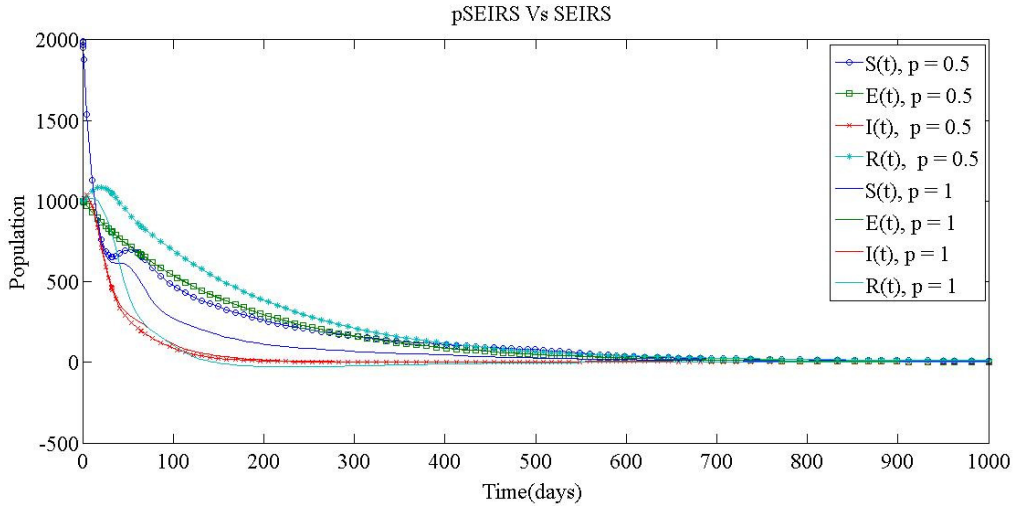

Fig. 17: Comparison between the classical SEIRS (where *p*=1) and the *p*SEIRS model.

In Table 5, we present a comparison of the model under different choices of the immunity probability *p*. The *p*SEIRS model gives a more realistic picture of the scale free network of 5000 nodes since it is unrealistic for all nodes to develop permanent immunity from getting infected with viruses, trojans and similar entities.

Table 5: Statistical Analysis of Fig. 16 and Fig. 17.

| Compartment | Probability | Min | Max | Mean |
|---|---|---|---|---|
| Susceptible | *p* = 0.25 | 8.008 | 2000 | 345.9 |
| | *p* = 0.5 | 5.379 | 2000 | 327.1 |
| | *p* = 0.75 | 8.41 | 2000 | 346.4 |
| | *p* = 1 | 5.797 | 2000 | 318.3 |
| Exposed | *p* = 0.25 | 2.509 | 1000 | 296.6 |
| | *p* = 0.5 | 2.515 | 1000 | 322 |
| | *p* = 0.75 | 2.504 | 1000 | 320.2 |
| | *p* = 1 | 2.506 | 1000 | 349.1 |
| Infected | *p* = 0.25 | 0.5927 | 1039 | 144.5 |
| | *p* = 0.5 | 1.799 | 1040 | 163 |
| | *p* = 0.75 | 1.212 | 1038 | 159.4 |
| | *p* = 1 | 3.047 | 1039 | 175 |
| Recovered | *p* = 0.25 | 8.63 | 1027 | 364.7 |
| | *p* = 0.5 | 3.497 | 1000 | 253.9 |
| | *p* = 0.75 | 9.491 | 1075 | 324.4 |
| | *p* = 1 | 0.154 | 1018 | 196.5 |

## 5. Conclusions and Discussion of Future Directions of Research

We have applied a variable population malicious object transmission model in complete and scale free networks, with constant latent and immune periods. The applied model extends the classical SIR model proposed in [11] to a *probabilistic SEIR* type model in several directions: (i) it includes an *Exposed* class in addition to the *Susceptible*, *Infected* and *Recovered*, (ii) it furthermore includes a constant exposition period $\omega$ and constant latency period $\tau$, (iii) when a node is removed from the infected class, it recovers with a temporary immunity with a probability *p* and

International Journal of Computer Networks & Communications (IJCNC) Vol.6, No.5, September 2014

dies due to the attack of the malicious object with probability (1-*p*). This is in contrast with Yan and Liu's model [16] which assumed that the node recovers with *permanent* immunity.

The model will be in an endemic state if the threshold parameter $R_0 > 1$. As $\omega$ increases, $R_0$ decreases and the condition of permanent infection will become less likely to be satisfied. Thus, the longer the exposure period of a system, the less likely it is that it will become endemic in the long run.

Another important information that we can gleam from the proposed model is the maximum number of nodes that will be infected. Knowing this number, we can assume that the most connected nodes are the most vulnerable to an attack; thus special attention should be paid to these nodes. This information is of great importance to network administrators. Moreover, one can get the global dynamic behavior of the network under certain assumptions, as illustrated in our simulations.

The SEIRS and other related compartment models are good tools to study the global behavior of an epidemic in a network or a population dynamic. They are relatively simple to implement, yet they can still give valuable information about the network dynamics. One of the major issues for these types of models is setting the right set of values for the parameters, thus making them highly restricted in terms of their behavior. To make these models more realistic, it is important to make them as robust as possible.

Some of the possible ways in which future work in this topic can expand is to consider a dynamic rate of change, i.e., dynamic death rate, transmission rate, and latency period. Demirci et al. [7] considered only one such parameter. Ozalp and Demirci [13] also considered a non-constant population within their SEIRS model. This is a good starting point to improve the existing SEIRS model [13]. Choosing a nonlinear incident rate in a SIERS model is also another direction. Though Ning and Junhong [6] have done some work on this, one can think of improving their work.

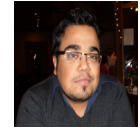

**Dr. Mohammad. S. Khan** received his M.Sc. and Ph.D. in Computer Science and Computer Engineering from the University of Louisville, Kentucky, USA, in 2011 and 2013 respectively. He is currently an Assistant Professor of Computer Science at Sullivan University. His primary area of search is in ad-hoc networks and network tomography. His research interest span several fields including ad-hoc network, mobile wireless mesh network and sensor network, statistical modeling, ODE, wavelets and ring theory. He has been on technical program committee of various international conferences and technical reviewer of various international journals in his field. He is member of IEEE.